\documentclass[12pt]{article}

\usepackage{graphicx,a4,epsfig,rotate,amsmath,amssymb}
\usepackage{latexsym,color,slashed,citesort}
\newcommand{\be}{\begin{equation}}
\newcommand{\ee}{\end{equation}}
\newcommand{\bea}{\begin{eqnarray}}
\newcommand{\eea}{\end{eqnarray}}
\newcommand{\eq}{\begin{eqnarray}}
\newcommand{\en}{\end{eqnarray}}

\newcommand{\ed}{\end{document}}
\newcommand{\bc}{\begin{center}}
\newcommand{\ec}{\end{center}}

\begin{document}

\thispagestyle{empty}

\begin{flushright}
{\footnotesize HISKP--TH--08/23, FZJ-IKP-TH--2008--22	}
\end{flushright}

\begin{center}

\vspace{2.0cm}
{\Large{\bf The mass of the \boldmath $\Delta$ resonance in a finite volume:\\[0.3em]
fourth-order calculation}}

\vspace{0.5cm}

\today

\vspace{0.5cm}

V. Bernard$^a$,
D. Hoja$^b$,
U.-G. Mei{\ss}ner$^{b,c}$ and
A.~Rusetsky$^b$

\vspace{2em}

\begin{tabular}{c}
$^a\,${\it 
Groupe de Physique Theorique,
IPN,
CNRS/Universit\'e Paris Sud 11}\\
{\it F-91406 Orsay Cedex France}\\[2mm]
$^b\,${\it Helmholtz--Institut f\"ur Strahlen-- und Kernphysik and}\\
{\it Bethe Center for Theoretical Physics}\\
{\it Universit\"at Bonn, D--53115 Bonn, Germany}\\[2mm]
$^c${\it  Institut f\"ur Kernphysik, Institute for Advanced Simulations}\\
{\it and J\"ulich Center for Hadron Physcis}\\ 
{\it Forschungszentrum J\"ulich, D-52425 J\"ulich, Germany}
\end{tabular}

\end{center}

\vspace{1cm}

{\abstract
{
We calculate the self-energy of the $\Delta (1232)$ resonance in a finite volume,
using chiral effective field theory with explicit spin-3/2 fields.
The calculations are performed up-to-and-including fourth order in the small
scale expansion and yield an explicit parameterization of the energy spectrum
of the interacting pion-nucleon pair in a finite box in terms of both the quark mass
and the box size $L$. It is shown that finite-volume corrections can be sizeable
at small quark masses.
}}

\vskip1cm

{\footnotesize{\begin{tabular}{ll}
{\bf{Pacs:}}$\!\!\!\!$& 11.10.St, 11.15.Ha, 12.39.Fe\\
{\bf{Keywords:}}$\!\!\!\!$& Resonances in lattice QCD,
field theory in a finite volume,\\$\!\!\!\!$ &chiral perturbation theory,
small scale expansion
\end{tabular}}
}
\clearpage


\section{Introduction}

The recent surge of interest in lattice calculations of  
the excited baryon
spectrum~\cite{Richards:2001bx,Maynard:2002ys,Gattringer:2003qx,Sasaki:2001nf,Zanotti:2001yb,Lasscock:2007ce,Melnitchouk:2002eg,Zhou:2006xe,Mathur:2003zf,Guadagnoli:2004wm,Alexandrou:2007qq,Alexandrou,McNeile:2003dy,Leinweber:2004it,Gattringer:2007da,Alexandrou:2008bn,Mathur:2008gw,Morningstar:2008mc,Fodor} 
has been mainly motivated by the experimental resonance physics program at Jefferson 
Lab~\cite{Jlab} and ELSA~\cite{BeckThoma}. Also, the hadron spectrum is
arguably the least understood feature of Quantum Chromodynamics.
In general, the extraction of the
properties of the excited states from the lattice data is a more delicate
enterprise as compared to the ground-state hadrons. The reason is that the
excited states are unstable and, strictly speaking, can not be put in 
correspondence to a single isolated level in the discrete spectrum measured
in lattice simulations. A standard procedure proposed by 
L\"uscher~\cite{LuescherII,Luescher_torus,Luescher_rho,Houches} 
(see also \cite{Wiese,DeGrand,Rummukainen:1995vs,Kim:2005gf,Christ:2005gi}) consists in placing the system into a finite 
cubic box of a size $L$ and studying the response of the spectrum on the change
of $L$. It can be shown that the dependence of the energy levels on $L$ is 
dictated solely by the scattering phase shift in the infinite volume. 
Consequently, the method is capable of extracting the phase shift from the 
lattice data that also
determines the position and the width of the resonances 
(see, e.g.~\cite{Aoki:2007rd,Gockeler:2008kc,Fodor}). Recently, the above
approach has been also applied to study nucleon-nucleon phase shifts at low
energy, as well as the two-body shallow bound 
states~\cite{Beane:2003yx,Beane:2003da,Beane:2006gf,Beane:2006mx,Sasaki:2006jn}.

Alternative approaches to study the decaying states have been
suggested, see, e.g.~\cite{Michael,Loft:1988sy,Lellouch:2000pv}. 
In particular, an interesting proposal is to reconstruct the spectral function
by using the maximal entropy method~\cite{Yamazaki}, which can also be
used to address the problem of unstable systems.

In actual calculations on the lattice the quark  masses do not usually
coincide with physical quark masses. This qualitatively changes the 
picture since, if the quark mass is large enough, the $\Delta (1232)$
does not decay and can be extracted by the methods applicable in case of 
the stable particles. Reducing the quark mass, 
a value is reached such that the  
$\Delta$ starts to decay into a pion and a nucleon\footnote{The decay
threshold is located at $M_N+M_\pi=M_\Delta$ in the infinite volume.
In a finite volume, the decay of $\Delta$ at threshold in the center-of-mass
(CM) 
frame is forbidden. Still, for brevity, the point $M_N+M_\pi=M_\Delta$ 
will be always referred below to as the threshold.}. The spectrum becomes strongly 
volume-dependent and L\"uscher's method has to be applied to extract
the parameters of the resonance -- the mass and the width.

Above the threshold $M_N+M_\pi>M_\Delta$, the finite-volume corrections to the 
spectrum are exponentially suppressed
and can be neglected in the first approximation. However, for those
values of the quark masses which correspond to
$M_N+M_\pi<M_\Delta$,
finite-volume corrections may become large and should be taken into account.
Note that merely making the volume larger does not suffice in the case
of an unstable state. Due to the potentially large corrections, 
the finite volume data
on the finite-volume energy spectrum 
can be enhanced below threshold. This enhancement, which is visible in the
lattice data at 
smaller volumes, can not be described by using the formulae for the
quark mass dependence in the infinite volume. 
We shall demonstrate an explicit example of such a behavior below.

From the above discussion it is clear that, in
order to be able to include all available lattice data for large as well 
as small quark masses in the analysis, one needs to provide a simultaneous 
explicit parameterization of the lattice QCD  spectrum in terms of both 
the quark mass $\hat m$ and the box size $L$.
This goal can be achieved by invoking the chiral 
effective field theory with explicit spin-3/2 degrees of 
freedom~\cite{Jenkins:1991es,Hemmert:1997ye} 
in a finite volume. The first attempt in this direction
was made in Ref.~\cite{delta1}, where we have performed the calculations
of the finite-volume energy spectrum at third order in the so-called small
scale expansion (SSE). The present paper extends these calculations to the
fourth order. In addition,
\begin{itemize}

\item[i)]
We provide an explicit 
formula for the finite-volume corrections for the unstable
$\Delta$, which can be used in the analysis of the lattice data;

\item[ii)]
We perform a fit of the obtained expressions to the most recent 
available data at different quark masses,
{\em taking into account finite-volume corrections}.
The fit allows one to determine some of the low-energy 
constants (LECs) in the chiral Lagrangian;

\item[iii)]
In doing so, one does not need to resort to any input 
phenomenological parameterization
of the resonant amplitude, because SSE provides such a parameterization
automatically, order by order in the $\epsilon$-expansion
(here, $\epsilon$ denotes the 
formal small expansion parameter in the SSE).

\item[iv)]
We analyze the quark mass dependence of the spectrum 
by using the method of probability distribution, introduced 
in~\cite{delta2}.

\end{itemize}

Note also that in this paper we do not consider the finite-volume effects
in the stable particle masses, which are exponentially suppressed at large 
volumes. Such effects can be treated within the same approach, see, 
e.g. Ref.~\cite{AliKhan:2003cu}.

The layout of the paper is as follows. In section~\ref{sec:infinite}
we discuss the calculation of the mass of the nucleon and the $\Delta$ in the
infinite volume, at fourth order in the small scale expansion.
In section~\ref{sec:finite} the calculation of the finite-volume energy 
spectrum of the $\pi N$ system is addressed. 
In section~\ref{sec:fit} we 
consider the fit of the explicit analytic expressions for the nucleon and
$\Delta$ mass to the existing data from lattice QCD and determine some
of the LECs of the chiral Lagrangian. We also analyze the finite
volume spectrum with the use of probability distributions~\cite{delta2}. 
Finally, section~\ref{sec:concl} contains our conclusions.

\section{The mass of the nucleon and the
$\Delta$ resonance in the infinite volume}
\label{sec:infinite}

Our calculations will be carried out in two steps. We first
perform calculations of the nucleon and $\Delta$ mass 
at order $\epsilon^4$ in the infinite volume\footnote{The
  small parameter $\epsilon$ subsumes  external momenta, the pion mass and
  the nucleon-delta mass splitting.} .
At the second step, we use the same Lagrangian
 in order to carry out the calculations
of the finite-volume energy spectrum. The results of these calculations
are applied to the case of an unstable $\Delta$.

The Lagrangian of pions, nucleons and deltas up-to-and-including order
$\epsilon^4$ in the SSE is taken from
Ref.~\cite{Bernard:2005fy}. Below we display only those
terms that contribute to the nucleon and $\Delta$ mass at this order,
\begin{equation}
 {\cal L}=\sum\limits_{i=1}^4\left( {\cal L}_{\pi N}^{(i)}
+{\cal L}_{\pi\Delta}^{(i)}\right )
+\sum\limits_{i=1}^2{\cal L}_{\pi N\Delta}^{(i)}\, ,
\end{equation}
where the pion--nucleon  Lagrangians are given by
\begin{eqnarray}
 {\cal L}_{\pi N}^{(1)}&=&
\bar{\psi}_N\left[ i\slashed{D}-\mathring{m}_N
+\frac{g_A}{2}\,\slashed{u}\gamma_5\right] \psi_N\, ,
\nonumber\\[2mm]
 {\cal L}_{\pi N}^{(2)}&=&\bar{\psi}_N\left[ c_1\langle\chi_+\rangle 
-\frac{c_2}{4\mathring{m}_N^2}\,\left( \langle u_\mu u_\nu\rangle D^\mu D^\nu 
+\text{h.c.}\right) +\frac{c_3}{2}\,\langle u^2\rangle +\ldots\right] \psi_N\, ,
\nonumber \\[2mm]
{\cal L}_{\pi N}^{(3)}&=&\bar{\psi}_N\left[B_{23}\Delta_0\langle\chi_+\rangle
+B_{32}\Delta_0^3+\ldots\right] \psi_N\, ,
\nonumber
\en
\eq
{\cal L}_{\pi N}^{(4)}&=&\bar{\psi}_N\Bigl[e_{38}\langle\chi_+\rangle^2
+\frac{1}{4}\,e_{115}\langle\chi_+^2-\chi_-^2\rangle 
-\frac{1}{4}\,e_{116}\left[ \langle\chi_-^2\rangle -\langle\chi_-\rangle^2
+\langle\chi_+^2\rangle-\langle\chi_+\rangle^2\right] 
\nonumber\\[2mm] 
&+&E_1\Delta_0^4+E_2\Delta_0^2\langle\chi_+\rangle+\ldots\Bigr] \psi_N\, ,
\end{eqnarray}
and
\begin{eqnarray}
{\cal L}_{\pi\Delta}^{(1)}&=&-\bar{\psi}_\alpha^iO^{\alpha\mu}\Bigl\lbrace
\left[i\slashed{D}^{ij}-\mathring{m}_\Delta\xi_{3/2}^{ij}
+\frac{g_1}{2}\,\slashed{u}^{ij}\gamma_5\right] g_{\mu\nu}
-\frac{1}{4}\left[\gamma_\mu\gamma_\nu\, , \,\left(i\slashed{D}^{ij}
-\mathring{m}_\Delta\xi_{3/2}^{ij}\right)\right]
\Bigr\rbrace\, O^{\nu\beta}\psi_\beta^j\, ,
\nonumber\\[2mm]
 {\cal L}_{\pi \Delta}^{(2)}&=&-\bar{\psi}_\alpha^iO^{\alpha\mu}\Bigl\lbrace 
\Bigl[  a_1\langle\chi_+\rangle\delta^{ij} 
-\frac{a_2}{4\mathring{m}_\Delta^2}\,\left( \langle u_\rho u_\sigma\rangle 
D_{ik}^\rho D_{kj}^\sigma +\text{h.c.}\right)
\nonumber\\[2mm]
&+& 
\frac{a_3}{2}\,\langle u^2\rangle\delta^{ij} +\ldots\Bigr] g_{\mu\nu}
+\ldots\Bigr\rbrace O^{\nu\beta} \psi_\beta^j\, ,
\nonumber \\[2mm]
{\cal L}_{\pi \Delta}^{(3)}&=&-\bar{\psi}_\alpha^iO^{\alpha\mu}
\left[B_1^\Delta\Delta_0\langle\chi_+\rangle+B_0^\Delta\Delta_0^3+\ldots\right]
g_{\mu\nu}\delta^{ij}O^{\nu\beta} \psi_\beta^j\, ,
\nonumber\\[2mm]
{\cal L}_{\pi \Delta}^{(4)}&=&-\bar{\psi}_\alpha^iO^{\alpha\mu}
\Bigl[e_{38}^\Delta\langle\chi_+\rangle^2
+\frac{1}{4}\,e_{115}^\Delta\langle\chi_+^2-\chi_-^2\rangle 
-\frac{1}{4}\,e_{116}^\Delta\left[ \langle\chi_-^2\rangle 
-\langle\chi_-\rangle^2
 +\langle\chi_+^2\rangle-\langle\chi_+\rangle^2\right]
\nonumber\\[2mm] 
&+&E_1^\Delta\Delta_0^4+E_2^\Delta\Delta_0^2\langle\chi_+\rangle+\ldots\Bigr] 
g_{\mu\nu}\delta^{ij}O^{\nu\beta} \psi_\beta^j\, .
\end{eqnarray}
The $\pi N\Delta$ interaction is described by the following Lagrangians
\begin{eqnarray}
{\cal L}_{\pi N\Delta}^{(1)}&=&c_A\bar{\psi}_\alpha^i O^{\alpha\beta}
w_\beta^i\psi_N\; +\; \text{h.c.}\, ,
\nonumber\\[2mm]
{\cal L}_{\pi N\Delta}^{(2)}&=&\bar{\psi}_\alpha^i O^{\alpha\mu}
\left[ib_3w_{\mu\nu}^i\gamma^\nu
+i\frac{b_6}{\mathring{m}_N}\,w_{\mu\nu}^iiD^\nu+\ldots\right] \psi_N
+\text{h.c.}\, .
\end{eqnarray}

\noindent
In the above expressions, $\psi_N$ and $\psi^i_\mu$ denote the nucleon and
the $\Delta$ field, respectively, $\mathring{m}_N$ and $\mathring{m}_\Delta$
stand for their masses in the chiral limit and $\Delta_0=\mathring{m}_\Delta
-\mathring{m}_N$. Note that $M_\pi=O(\epsilon)$ and $\Delta_0=O(\epsilon)$.
The building blocks that are used in the construction of the
above Lagrangian are given by
\begin{eqnarray}
 &&U=u^2\, , \qquad u_\mu=iu^\dagger\partial_\mu U u^\dagger\, ,
\qquad D_\mu=\partial_\mu+\frac{1}{2}[u^\dagger,\partial_\mu u]~,
\nonumber\\[2mm]
&&\chi=2B(s+ip)\, , \qquad 
\chi_\pm=u^\dagger\chi u^\dagger\pm u\chi^\dagger u\, , \qquad 
s=\hat{m}\textbf{1}+\ldots ~,
\nonumber\\[2mm]
&&D_{ij}^\mu=\delta_{ij}D^\mu-i\epsilon_{ijk}\langle \tau^kD^\mu\rangle\, , 
\qquad 
u_{ij}=\delta_{ij}u^\mu\, , \qquad 
w_\mu^i=\frac{1}{2}\langle\tau^iu_\mu\rangle~.
\nonumber \\[2mm]
&&w^i_{\mu\nu}=\langle\tau^i[D_\mu,u_\nu]\rangle/2\, ,\qquad 
O^{\mu\nu}=g^{\mu\nu}-\frac{2}{d}\,\gamma^\mu\gamma^\nu\, ,
\end{eqnarray}
and the isospin projectors are defined by
\begin{equation}
\xi_{ij}^{3/2}=\delta_{ij}-\frac{1}{3}\tau_i\tau_j\, ,\qquad \xi_{ij}^{1/2}=\frac{1}{3}\tau_i\tau_j\,.
\end{equation}
In these formulae standard notation is utilized. 
Namely, we use $U=\exp(i\mathbf{\tau}\cdot\mathbf{\pi}/F)$, where
$\pi$ is the pion field. We work in the isospin limit $m_u=m_d=\hat{m}$ 
and the trace in flavor space is denoted by $\langle \ldots\rangle$. 
The quantity $F$ is the pion decay constant, $B$ is related to the quark 
condensate and $g_A$ is
the nucleon axial-vector constant (all in the chiral limit). 
 The coefficients $c_i,a_i,\cdots$ are the pertinent LECs.

The propagator of a Rarita-Schwinger field in $d$ dimensions is given by
\begin{equation}
S_{\mu\nu}^{(0)}=-\frac{1}{\mathring{m}_\Delta-\slashed{p}}
\left[ g_{\mu\nu}-\frac{1}{d-1}\gamma_\mu\gamma_\nu
-\frac{d-2}{(d-1)(\mathring{m}_\Delta)^2}p_\mu p_\nu
+\frac{p_\mu\gamma_\nu-p_\nu\gamma_\mu}{(d-1)\mathring{m}_\Delta}\right]\,\xi_{ij}^{3/2}\, . 
\end{equation}

\noindent
The calculations are carried out in infrared regularization. Differently
from Ref.~\cite{Hemmert,Bernard:2005fy}, we do not
project out the redundant spin-$1/2$ components of the $\Delta$ propagator,
which appears in the loops. This amounts merely to a redefinition of some of the
LECs -- hence, the numerical values of LECs determined from fitting
to the same data, should in general differ in these two schemes. For related 
discussion of this issue, see also \cite{Tang:1996sq,Ellis:1996bd,Fettes:2000bb,Bernard-Progress,Krebs:2007rh}.

\begin{figure}[t]
\centerline{
\includegraphics[width=12cm]{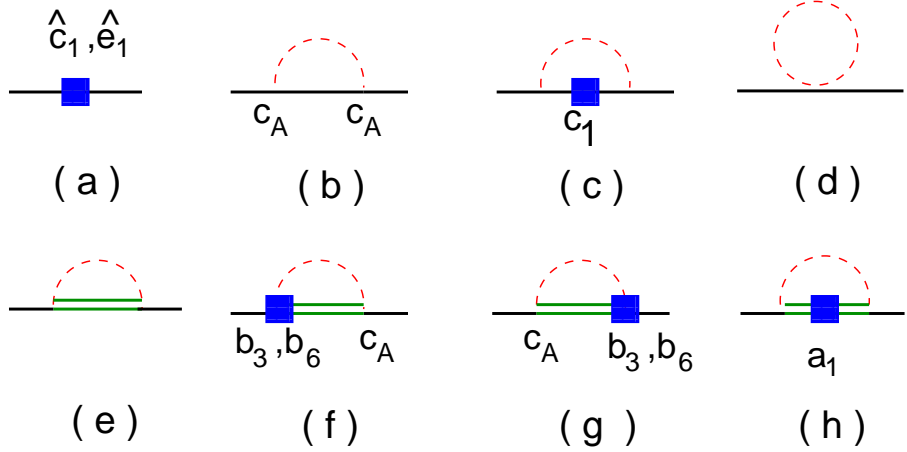}}
\caption{Graphs contributing to the nucleon self-energy at 
${\cal O}(\epsilon^4)$ in SSE. Solid, double solid and dashed lines denote
nucleons, deltas and pions, in order.}
\label{nucleongraphs}
\end{figure}

\noindent

The self-energy of the $\Delta$ is complex on the mass shell for those
values of the pion masses, when the $\Delta$ turns unstable, i.e.,
$M_\pi<M_\Delta-M_N$. The mass of the 
$\Delta$ is defined as a real part of the pole position in the propagator.

The diagrams that contribute to the nucleon and $\Delta$ masses at order
$\epsilon^4$, are displayed in Fig.~\ref{nucleongraphs} and 
Fig.~\ref{deltagraphs}, respectively. The calculations are pretty standard
and the final results are listed in Appendix~\ref{app:masses}. 
Since we are primarily interested in fitting the quark mass dependence
to lattice data,
it is useful to normalize both quantities at the physical value of the
quark (pion) mass
\eq\label{eq:fitformulae}
M_N\!\!&=&\!\!\bar M_N+x_1(M_\pi^2-\bar M_\pi^2)+x_2(M_\pi^3-\bar M_\pi^3)
+x_3(M_\pi^4-\bar M_\pi^4)
+x_4\biggl(M_\pi^4\ln\frac{M_\pi}{m_N}-\bar M_\pi^4\ln\frac{\bar M_\pi}
{\bar M_N}\biggr)
\nonumber\\[2mm]
&-&\frac{Z}{F^2}\,\bigl(\Phi_N(m_N,m_\Delta,M_\pi^2)
-\Phi_N(\bar M_N,\bar M_\Delta,\bar M_\pi^2)\bigr)+O(\epsilon^5)\, ,
\nonumber\\[2mm]
M_\Delta\!\!&=&\!\!\bar M_\Delta+y_1(M_\pi^2-\bar M_\pi^2)
+y_2(M_\pi^3-\bar M_\pi^3)+y_3(M_\pi^4-\bar M_\pi^4)
+y_4\biggl(M_\pi^4\ln\frac{M_\pi}{m_N}-\bar M_\pi^4\ln\frac{\bar M_\pi}{\bar M_N}\biggr)
\nonumber\\[2mm]
&-&\frac{Z}{F^2}\,\bigl(\Phi_\Delta(m_N,m_\Delta,M_\pi^2)
-\Phi_\Delta(\bar M_N,\bar M_\Delta,\bar M_\pi^2)\bigr)+O(\epsilon^5)\, ,
\en
where in the fit we use
\eq\label{eq:NDmasses}
m_N&=&\bar M_N+x_1(M_\pi^2-\bar M_\pi^2)+\cdots\, ,\quad\quad
m_\Delta=\bar M_\Delta+y_1(M_\pi^2-\bar M_\pi^2)+\cdots\, ,
\nonumber\\[2mm]
Z&=&c_A^2 +2(m_\Delta-m_N)c_Ab_3
+\frac{m_\Delta^2-m_N^2-M^2}{m_N}\,c_Ab_6
\nonumber\\[2mm]
&=&c_A^2+2\Delta_0\,c_A(b_3+b_6)+\cdots\, .
\en
Here, $M^2=2\hat mB$ and $\bar M_\pi,\bar M_N,\bar M_\Delta$ stand for the physical values
of the pion, nucleon and $\Delta$ masses. The ellipses denote the higher-order
terms in $\epsilon$.  Further, at the  order we are working, 
one may take $\Delta_0=\bar M_\Delta-\bar M_N+\cdots$ in the above equations. The masses
$M_N,M_\Delta$ are  functions of the pion mass $M_\pi$. The quantities
$x_i,y_i,Z$ denote certain combinations of LECs. Explicit expressions
for  the $x_i,y_i,Z$, as
well as for the functions $\Phi_N,\Phi_\Delta$
are displayed in Appendix~\ref{app:masses}.
Fitting the nucleon and $\Delta$ masses, given by Eq.~(\ref{eq:fitformulae}),
to the lattice data determines the numerical values of the above 
combinations of LECs. Note that some higher-order terms are also present in
Eq.~(\ref{eq:NDmasses}), e.g. in the expressions for $\Phi_\Delta,\Phi_N$.

\begin{figure}[t]
\centerline{
\includegraphics[width=12cm]{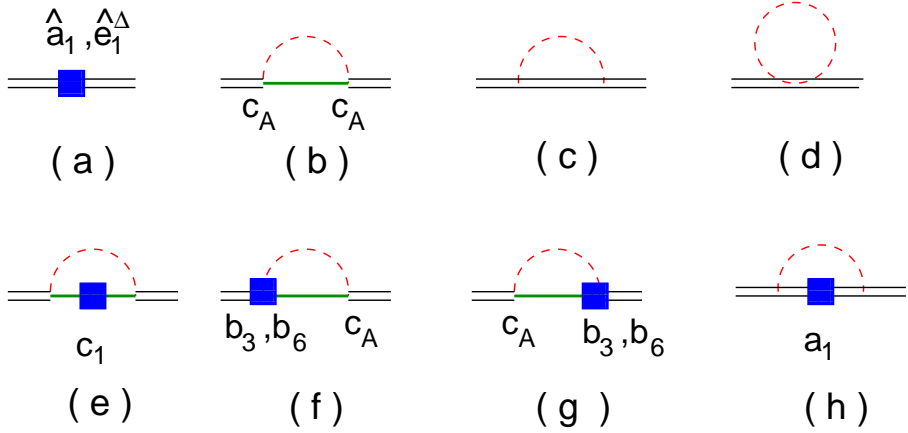}}
\caption{Graphs contributing to the self-energy of $\Delta$ at 
${\cal O}(\epsilon^4)$ in SSE. For notation, see Fig.~\ref{nucleongraphs}.}
\label{deltagraphs}
\end{figure}

\noindent
The calculation of the quark mass dependence of the nucleon and $\Delta$ masses
has been carried out in different 
settings~\cite{ProcuraHemmert,Tiburzi,Hemmert,Bernard:2005fy,Pascalutsa}.
Note that, in particular,
 our result for the nucleon mass in the infinite volume 
agrees at $O(\epsilon^3)$ with the expression given in 
Eq.~(17) of Ref.~\cite{ProcuraHemmert}.
However, it differs from the $O(\epsilon^4)$ result 
for the nucleon and $\Delta$ masses, which are displayed in Eqs.~(22) and (30)
of Ref.~\cite{Tiburzi}, respectively. For instance, these latter expressions
do not contain the LECs which describe the quark mass dependence of the 
$\pi N\Delta$ vertex (analog of the constants $b_3,b_6$).

\section{Self-energy of the $\Delta$ resonance in a finite volume: 
the energy levels}
\label{sec:finite}

\subsection{Calculation of the finite-volume correction}

In a finite volume the $\Delta$ propagator develops a tower of poles
on the real axis. The location of these poles determines the finite-volume
energy spectrum of the system. Thus, calculating the propagator in a
finite volume, we shall be able to study the volume-dependence of the
energy levels. The procedure is described in detail in Ref.~\cite{delta1} and
will not be repeated here.
Here we simply note that the only difference to the infinite-volume case
is the replacement of the (Euclidean) loop integrations by infinite
sums
\eq
\int\frac{d^4 k_E}{(2\pi)^4}\,(\cdots)\mapsto\int \frac{dk_4}{2\pi}\,
\frac{1}{L^3}\,\sum_{\bf k}\,(\cdots)\, ,
\qquad
{\bf k}=\frac{2\pi}{L}\,{\bf n}\, ,\quad{\bf n}\in \mathbb{Z}^3\, .
\en
In the above expression,
$L$ denotes the size of the (cubic) box in which the system is placed.
The Lagrangian that produces the loops is the same as in the infinite volume.

The calculations are substantially simplified, if carried out in a large
volume where the exponentially suppressed corrections
can be neglected. In this limit the masses of the stable particles 
can be considered as volume-independent.
However, as it is well known, the energy levels corresponding to the unstable
particles receive corrections, which are suppressed by powers
of $L$. Only the diagrams, which contain the pion-nucleon intermediate
state -- diagrams b,e,f,g in Fig.~\ref{deltagraphs}, calculated
in a finite volume -- contribute to this power-like behavior. Retaining 
the finite-volume parts of these diagrams only,
the equation that determines the 
location of the poles in the $\Delta$ propagator is written as
(cf with Ref~\cite{delta1})
\eq
\label{eq:central}
 M_\Delta-E=\frac{\tilde{Z}}{2EF^2}\,
\left( (E+M_N)^2-M_\pi^2 \right)
\frac{\lambda(E^2,M_N^2,M_\pi^2)}{12E^2}\,\tilde{W}_0^N(E^2), .
\en
Here, $E$ denotes the pole position on the real axis, and $M_\pi,M_N,M_\Delta$
are the masses in the infinite volume. Further,
$\tilde{Z}$ stands for the following combination of the LECs
\eq\label{eq:tildeZ}
 \tilde{Z}&=&c_A^2+2b_3c_A(E-M_N)+2b_6c_A\,\frac{E^2-M_N^2-M_\pi^2}{2M_N} 
\nonumber\\[2mm]
&=&Z+2(b_3+b_6)c_A(E-M_\Delta)+{\cal O}(\epsilon^2)\, .
\en
It is seen that only three LECs: $c_A,b_3,b_6$ appear in the finite-volume
correction to the energy of $\Delta$.

Finally, the quantity $\tilde W_0^N(E^2)$ corresponds to the finite-volume
part of the $\pi N$ loop
function
\eq
\label{umformung}
 \tilde{W}_0^N(E^2)=W_0^N(E^2)-W_0^N(E^2)\Big|_{L\rightarrow\infty},
\en
where
\eq
W_0^N(E^2)=\int\frac{dk_4}{2\pi}\frac{1}{L^3}\sum\limits_{\vec k}
\frac{1}{(M_\pi^2+k^2)(M_N^2+(\hat{P}-k)^2)}\, ,\qquad 
\hat P_\mu=(iE,{\bf 0})\, .
\en
In  large volumes, 
neglecting exponentially suppressed contributions,
 the loop function above threshold can be 
rewritten as
\eq\label{eq:largeL}
\tilde{W}_0^N(E^2)&=&\frac{1}{4\pi^{3/2}\,EL}\,\bar {\cal Z}_{00}(1,q^2)+\cdots
\, ,
\nonumber\\[2mm]
\bar {\cal Z}_{00}(1,q^2)&=&{\cal Z}_{00}(1,q^2)
-{\cal Z}_{00}(1,q^2)\biggr|_{L\to\infty}\, ,
\en
where the ellipses stand for the exponentially suppressed contributions,
the quantity $q=\frac{L}{2\pi}\, p$ with 
$p=\lambda^{1/2}(E^2,M_N^2,M_\pi^2)/2E$ and 
${\cal Z}_{00}$ is the zeta-function from Ref.~\cite{Luescher_torus}
\begin{equation}\label{eq:defzeta}
{\cal Z}_{00}(s,q^2)=
\frac{1}{\sqrt{4\pi}}\sum\limits_{{\bf n}\in\mathbb{Z}^3}
\frac{1}{({\bf n}^2-q^2)^{s}}\, .
\end{equation}
Note that $\bar {\cal Z}_{00}(1,q^2)={\cal Z}_{00}(1,q^2)$ for $q^2>0$.

\subsection{Relation to L\"uscher's formula}

\begin{figure}[t]
\centering
 \includegraphics[width=6cm]{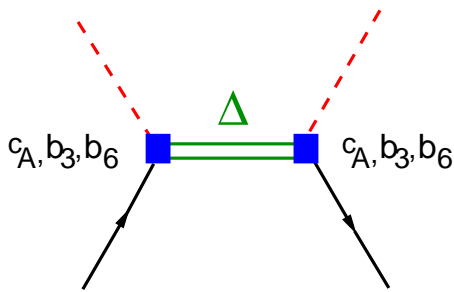}
\caption{Feynman diagram yielding the same 
scattering phase as Eq.~(\ref{tandelta}) from an infinite volume SSE
calculation. Solid, dashed and double lines denote nucleons, pions and deltas,
respectively.
}
\label{scattering}
\end{figure}

By using Eqs.~(\ref{eq:largeL}) and (\ref{eq:defzeta}) 
it can be checked that the Eq.~(\ref{eq:central}) which determines the
position of the pole in the propagator, can be rewritten 
in the form of L\"uscher's equation
\begin{equation}
\label{zeta}
 \tan\delta(p)=\frac{\pi^{3/2}q}{{\cal Z}_{00}(1,q^2)}\, ,
\end{equation}
where $\delta(p)$ denotes the scattering phase shift in the $P_{33}$-channel
 for the following choice of the scattering phase
\begin{equation}
\label{tandelta}
\tan\delta(p)=\frac{p^3}{48\pi E^2}\cdot\frac{(E+M_N)^2-M_\pi^2}{M_\Delta-E}
\cdot\frac{\tilde{Z}}{F^2}\, ,
\end{equation}
which corresponds to the $s$-channel tree-level scattering amplitude in the 
SSE, shown in Fig.~\ref{scattering}.
The discrete solutions of Eq.~(\ref{zeta}) determine the energy
spectrum of the system $E_n=\sqrt{M_N^2+p_n^2}\\+\sqrt{M_\pi^2+p_n^2}$
through the given scattering phase $\delta(p)$.

\subsection{Effect due to the finite lattice spacing}

Certain caution is needed, if one uses the above 
formulae in order to fit the lattice data. Indeed, they
 contain artefacts
due to the finite lattice spacing $a$. For example, in the analysis of the
data obtained by using twisted mass fermions, one has to address the issue
of isospin breaking at  finite $a$. Even if the effect turns out to be
not very large  in the measured nucleon and delta masses, 
the neutral pion masses in
the loops will differ strongly from the charged ones. It is clear
that, in order to address the problem in its full generality, one has to
develop twisted mass chiral perturbation theory, where the isospin breaking
emerges at a finite lattice spacing. In this paper, however, we shall restrict
ourselves to the spectrum of $\Delta^{++},\Delta^-$, where only charged pions
occur in the loops up-to-and-including order $\epsilon^4$. Consequently, at
this order one may use the conventional formalism,
with the pion mass set equal to the charged pion mass and assuming 
isospin symmetry in the couplings.
The data on $\Delta^+,\Delta^0$ will be used for checking the size
of isospin-breaking contributions at  finite $a$ and thus will
serve as an error 
estimate only.

\subsection{Determination of the width}

The width of the $\Delta$ at the physical value of the quark mass
is determined by the parameter $Z$ which, in turn,
at this order depends on the LECs $c_A,b_3,b_6$, see Eq.~(\ref{eq:NDmasses})
\eq\label{eq:Gamma}
\Gamma_\Delta=\frac{Zq_{cm}^3}{6\pi F_\pi^2}\,
\frac{(\bar M_\Delta+\bar  M_N)^2-\bar M_\pi^2}{4\bar M_\Delta^2}\, .
\en
In the above equation,
 $q_{cm}$ denotes the CM momentum of the $\pi N$ pair after the decay
of $\Delta$, $F_\pi$ is the pion decay constant and 
$\bar M_\Delta=1232~\mbox{MeV}$.

A determination of the LECs $c_A,b_3,b_6$ from the fit to the $\Delta$ mass 
in the infinite volume does not provide sufficient accuracy, 
because these LECs enter starting from the
 next-to-leading order. The situation changes, however, if we consider
the data obtained at the same quark mass and at different volumes.
Consider, for instance, the data taken at two different values of $L$.
Since the mass of the $\Delta$ in the infinite
volume is, by definition, volume-independent, the following consistency
condition must hold at this order 
\eq\label{eq:condition}
M_\Delta=E_\Delta(L_1)+\delta E_\Delta(L_1,c_A,b_3,b_6)
=E_\Delta(L_2)+\delta E_\Delta(L_2,c_A,b_3,b_6)\, ,
\en
where $E_\Delta(L_i),~i=1,2$, denote the measured energies
and $\delta E_\Delta(L_i,c_A,b_3,b_6)$ denotes the
finite-volume correction below threshold evaluated at the pertinent values
of $E$ and $L$, see  Eq.~(\ref{eq:central}). Performing measurements at different
values of $L$ provides additional constraints.
Extracting the values of LECs from the
above conditions, one may in principle determine the width of the $\Delta$
by using Eq.~(\ref{eq:Gamma}). 
Note that Eq.~(\ref{eq:condition}) holds at
a fixed value of the quark mass.

\section{Fit to the lattice data}
\label{sec:fit}

\subsection{Choice of the data}

Just 
in order to demonstrate the application of the theoretical framework developed
above, we shall perform the fit to the recent data of the ETM 
collaboration~\cite{Alexandrou}. In particular, we fit the data for
the nucleon and $\Delta$ masses, obtained on $\beta=3.9$ lattices of size
$24^3\times 48$ and $32^3\times 64$ (smeared link and smeared source), 
corresponding to $L=2.1~\mbox{fm}$ and $L=2.7~\mbox{fm}$, respectively. 
These data are given in Table~II of Ref.~\cite{Alexandrou}. The data 
contain the
nucleon and $\Delta$ masses at 4 different values of the quark mass (on a 
smaller lattice) and one
 additional data point for the lightest quark mass
(on a larger lattice). At the lightest quark mass, the sum of the nucleon
and pion masses is smaller that the $\Delta$ mass.
Note that we do not have access to
the data at different volumes, extrapolated to the continuum limit $a\to 0$.
The values of the nucleon and $\Delta$ masses, displayed 
in Table~II of Ref.~\cite{Alexandrou} still contain the artefacts due to
a finite lattice spacing.

\subsection{Fit to the nucleon and $\Delta$ masses: infinite volume}

\begin{figure}[t]
\centering
 \includegraphics[width=10cm]{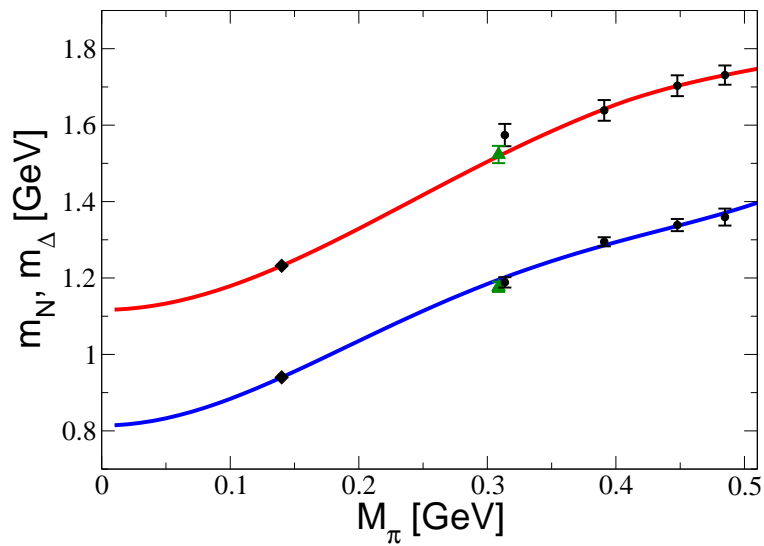}
\caption{The fit to the nucleon and $\Delta^{++}$ spectrum 
by using Eq.~(\ref{eq:fitformulae}). The filled circles
correspond to the data taken at $L=2.1~\mbox{fm}$.
The data corresponding to $L=2.7~\mbox{fm}$ at the smallest pion mass
are shown for comparison (triangles).
The black diamonds without error bars correspond to the physical masses.}
\label{fig:infinitevolume}
\end{figure}

In Ref.~\cite{Alexandrou} the infinite-volume mass of the $\Delta$ is identified
with the extracted energy level at a largest volume at a given quark mass.
As already mentioned, such a procedure can not be strictly 
justified for unstable
particles. Notwithstanding, we shall use this method in the beginning and
try to simultaneously fit both nucleon and $\Delta$ masses with the 
infinite-volume formulae~(\ref{eq:fitformulae}). The result is shown in 
Fig.~\ref{fig:infinitevolume}. For comparison,
in the same figure we display the data points taken on a smaller lattice.

The Eqs.~(\ref{eq:fitformulae}) contain too many free LECs, making
the fit to the few available data points questionable. A reasonable strategy
consists in constraining some of these LECs by using additional 
physical information. Thus, the LEC $x_2$ is unambiguously fixed through 
the known value of the nucleon axial-vector coupling $g_A=1.267$. 
Furthermore, we use the $SU(6)$-relation $g_1=(9/5) g_A$ and set $a_{2,3}=c_{2,3}$.
The LECs $c_{2,3}$ are determined by matching to ChPT without
explicit $\Delta$ degree of freedom. The pertinent relations are given
by $c_2=\tilde c_2-g_A^2/(2\Delta_0)+O(1)$ and 
$c_3=\tilde c_3+g_A^2/(2\Delta_0)+O(1)$, where $\tilde c_{2,3}$ denote the
LECs in ChPT without $\Delta$ (cf with Ref.~ \cite{Krebs:2007rh}).
 Using  the values 
$\tilde c_2=3.3~\mbox{GeV}^{-1}$ and 
$\tilde c_3=-4.7~\mbox{GeV}^{-1}$~\cite{Bernard-Progress}, we finally get
$c_2\simeq 0.55~\mbox{GeV}^{-1}$ and $c_3\simeq -1.95~\mbox{GeV}^{-1}$.
In addition, we use the value $Z=2.14$ that leads to the physical decay width
$\Gamma=118~\mbox{MeV}$ after substituting into Eq.~(\ref{eq:Gamma}).
The couplings $c_A$ and $b_3+b_6$ are given below, see Eq.~(\ref{eq:cAb36}).

The remaining LECs $\hat c_1, e_1,\hat a_1, e_1^\Delta$ 
are allowed to vary freely (these LECs are defined in Eq.~(\ref{eq:A3})). 
In the fit we will use the data at $L=2.1~\mbox{fm}$
except the lowest point corresponding to $L=2.7~\mbox{fm}$.
The fit to the data 
gives the following values for these parameters (no errors assigned)
\eq\label{eq:LECs-infinite}
&&\hat c_1=-1.6~\mbox{GeV}^{-1}\, ,\quad
\hat a_1=-1.8~\mbox{GeV}^{-1}\, ,
\nonumber\\[2mm]
&& e_1=-1.1~\mbox{GeV}^{-3}\, ,\quad
 e_1^\Delta=6.6~\mbox{GeV}^{-3}\, .
\en
The $SU(6)$ relation $\hat a_1\simeq \hat c_1$ holds approximately, in
difference with the result obtained in Ref.~\cite{Bernard:2005fy}
(note, however, that the different prescriptions for performing the infrared
regularization in the case of $\Delta$ amount to a finite renormalization of
various LECs). In order to compare the obtained value of $\hat c_1$ with
the phenomenological estimates, one has again to perform the matching to
ChPT without an explicit $\Delta$, which yields 
$\hat c_1=\tilde c_1+Z\Delta_0/(8\pi^2F^2)\,\ln(2\Delta_0/\bar m_N)+O(\Delta_0^2)$. 
In this expression, $\tilde c_1$ denotes the value of the pertinent LEC in ChPT.
The resulting shift $\simeq 0.44~\mbox{GeV}^{-1}$ in $\hat c_1$ 
is positive, and the obtained value of $\tilde c_1$ reasonably agrees 
with the value extracted
from the phenomenological analysis of the pion-nucleon scattering at fourth
order, see e.g.~\cite{physrep} for the latest update.
Note, however, that the fourth
order LECs $ e_1$ and $ e_1^\Delta$ differ significantly. Moreover, 
$ e_1^\Delta$ is rather large 
that could serve an indication of a poor convergence at these
pion masses.

As one observes from Fig~\ref{fig:infinitevolume}, 
the finite-size corrections to the 
$\Delta$ energy may turn out sizable below threshold (the data point 
corresponding to the smallest pion mass). At present, the error bars
on the data are large that precludes one to make an unambiguous statement
on the issue. However, even at the present accuracy a hint is seen that the
lowest data point at $L=2.1~\mbox{fm}$ is located above the curve. This is
an example of the enhancement which was mentioned in the introduction.

\subsection{Analytic behavior at threshold}

\begin{figure}[t]
\centering
 \includegraphics[width=8cm]{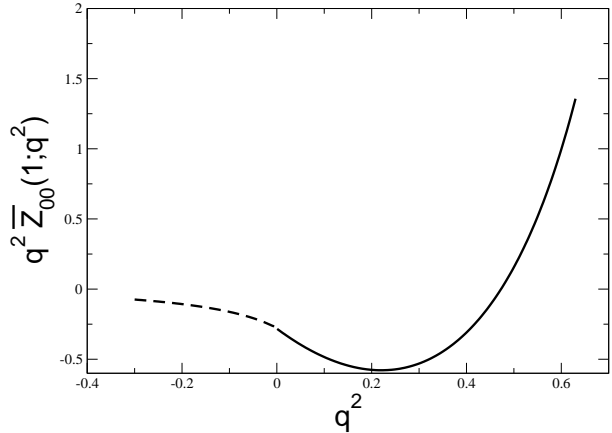}
\caption{The function $q^2\bar Z_{00}(1;q^2)$ in the vicinity of
threshold $q^2=0$. The threshold cusp is clearly visible.}
\label{fig:z00_plot}
\end{figure}

It is quite instructive to study the qualitative behavior of the energy levels
in the vicinity of threshold, i.e. choosing the quark mass so that
the sum of the pion and nucleon masses  are only
slightly below the $\Delta$ mass. As we know, this situation is realized
for the lowest data point.

Let us consider the plot
of the function $q^2\bar {\cal Z}_{00}(1;q^2)$, which enters the r.h.s. of 
Eq.~(\ref{eq:central}), see Fig.~\ref{fig:z00_plot}. 
This quantity has a {\em cusp}, proportional to $q^3$,
 at threshold $q^2=0$. Moreover,
its value in the limit $q^2\to 0$ is different from zero. Below threshold,
the function decreases exponentially. Above threshold, the function has a 
tower of poles, with the first one located at $q^2= 1$.

If one is varying the quark mass so that $q^2$ stays negative 
($\Delta$ stable), the finite-volume corrections are exponentially small.
However, if decreasing the quark mass, the quantity $q^2$ moves across the
cusp from below, the effect blows up rapidly. In this case, the energy
levels in a finite volume receive large
corrections, which should be taken into account.
On the other hand,  the ``raw'' data on the energy levels at a fixed 
volume, which are depicted, e.g. in Fig.~\ref{fig:infinitevolume},
are smooth functions of the quark mass
and do not exhibit any cusp.

\subsection{Subtracting finite-volume effect}

In order to subtract finite-volume effect at order $\epsilon^4$, one has to fix
the values of the LECs $c_A$ and $b_3+b_6$. Since we have only one data point
below threshold, both LECs can not be fixed simultaneously. 
For this reason, we have set the constant
$Z=2.14$ so as to reproduce the width of the $\Delta$
and used 
the consistency condition (\ref{eq:condition}) to determine $\tilde Z$ 
and thus to disentangle $c_A$ and
$b_3+b_6$ from Eqs.~(\ref{eq:tildeZ}) and (\ref{eq:defZ}). Using central values for the energy levels, we get
\eq\label{eq:cAb36}
c_A^2=2.73\, ,\quad  b_3+b_6=-0.6~\mbox{GeV}^{-1}\, .
\en
As seen, these LECs are indeed of the natural size.
 
\begin{table}[t]

\vspace*{.5cm}

\begin{center}

\begin{tabular}{|c|c|c|c|c|c|c|}

\hline
$L~\mbox{[fm]}$ & $M_\pi$ & $M_N$ & $E_{\Delta^{++,-}}$ & $E_{\Delta^{+,0}}$ 
& $\delta E_{\Delta^{++,-}}$ & $\delta E_{\Delta^{+,0}}$  \\
\hline\hline
2.1 & $314\pm 2.4$ & $1189\pm 14$ & $1574\pm 29$ & $1609\pm 40$  & -90 & -129 \\
\hline
2.7 & $309\pm 1.9$ & $1177\pm 13$ & $1523\pm 23$ & $1523\pm 34$  & -39 & -43 \\
\hline

\end{tabular}

\end{center}

\caption{Meson and baryon masses for two different values of the
box size $L$ (the data are taken from Table II of Ref.~\cite{Alexandrou} (central values only)
and correspond to the choice SS of the interpolating field). Last two columns
correspond to the finite-volume corrections to the energy levels,
calculated by using Eq.~(\ref{eq:central}) (see the text for more detail). 
All masses are given in MeV.}
\label{tab:hmasses}

\end{table}

\begin{figure}[t]

\vspace*{.2cm}

\centering
 \includegraphics[width=10cm]{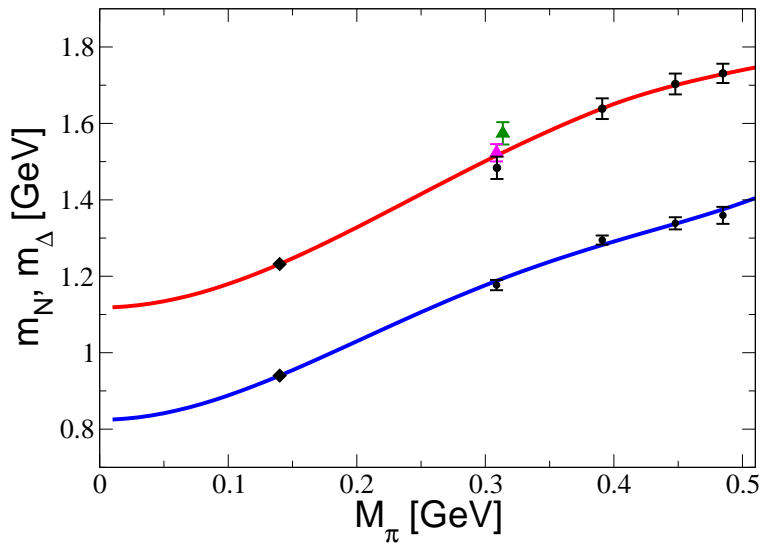}
\caption{The fit to the nucleon and $\Delta^{++}$ spectrum.
The lowest data point for $\Delta$ has been purified with respect to
the finite-volume corrections. For comparison, the uncorrected lowest data 
points for $L=2.1~\mbox{fm}$ and $L=2.7~\mbox{fm}$ (triangles) are shown.}
\label{fig:finitevolume}
\end{figure}

In Table~\ref{tab:hmasses} we give the results for the finite-volume 
corrections to the central value of the lowest data point, evaluated at the above values of
the LECs. These 
finite volume corrections are indeed small except for the lowest
point. 
The results for $\Delta^{+,0}$ are presented just for the 
visualization of the artefacts due to the finite lattice size. As is seen from 
this table, the finite-volume corrections matter even at the present accuracy.
For instance, the infinite-volume mass of the $\Delta^{++}$ is equal to
$1484~\mbox{MeV}$. Here we note that in Ref.~\cite{Walker-Loud} significant 
finite-volume corrections have been found as well. 
The calculations in  Ref.~\cite{Walker-Loud} have been carried out
at order $\epsilon^3$, by using the formula of Ref.~\cite{delta1}. 
At this order, one would set $c_A^2=Z=2.14$ and $b_3+b_6=0$ in our formulae.
It can be checked that this does not change the result significantly.

In Fig.~\ref{fig:finitevolume} we show the fit to the lattice data.
The finite-volume effect, which is given in Table~\ref{tab:hmasses}, 
is subtracted from the lowest data point. It can be seen that the LECs, which
are extracted from the fit, are quite stable (to be compared to Eq.~(\ref{eq:LECs-infinite}))
\eq\label{eq:LECs-finite}
&&\hat c_1=-1.6~\mbox{GeV}^{-1}\, ,\quad
\hat a_1=-1.7~\mbox{GeV}^{-1}\, ,
\nonumber\\[2mm]
&& e_1=-1.4~\mbox{GeV}^{-3}\, ,\quad
 e_1^\Delta=6.4~\mbox{GeV}^{-3}\, ,
\en
however, $\chi^2$ is somewhat worse in this case.

As can be observed from Fig.~\ref{fig:finitevolume}, the finite-volume
correction to the lowest data point is significant. There is no enhancement 
in the corrected data.

Finally, just as a hint,
we would like to mention that it is possible to get a very good
fit to the data, concentrating only on two lowest quark mass
data points and relaxing the condition $g_1=(9/5) g_A$. 
The obtained values for the LECs are $g_1=2.89\simeq 2.3g_A,~
\hat c_1 =-1.43~\mbox{GeV}^{-1}, \hat a_1 =-1.67~\mbox{GeV}^{-1}, 
 e_1= -1.35~\mbox{GeV}^{-3}$ and $ e_1^\Delta =2.02~\mbox{GeV}^{-3}$.
As can be seen, $ e_1^\Delta$ is now of natural size.
The constant $\hat c_1$, contributing to the nucleon $\sigma$-term at 
lowest order,
 turns out to be  slightly smaller. Of course, two data points do not provide
 sufficient input to draw definite conclusions about the values of
the LECs. For the same reason, we refrain here from citing the values of
the nucleon and delta $\sigma$-terms, which can be reliably determined, 
only if more data points become available at smaller quark masses.

\subsection{Probability distribution: ~dependence on the  quark mass}
\label{sec:probability}

In this section we shall study the quark (pion) mass dependence of the
structure of the energy levels. To this end, it is useful to invoke the 
language of the probability distributions~\cite{delta2}, which 
makes this dependence very transparent.

The probability distribution, which can be constructed from the 
volume-dependent energy spectrum through an unambiguous 
procedure~\cite{delta2}, is closely related to the so-called density of states
in a finite volume. Using L\"uscher's formula, it can be shown~\cite{delta2} 
that -- to a good approximation -- the probability distribution $W(p)$ can be
expressed via the scattering phase
\eq\label{eq:Wp}
W(p)=\frac{C}{p}\,\sum_{n=1}^N\biggl(\frac{\sqrt{4\pi(\pi n-\delta(p))}}{p}
+\frac{2\pi\delta'(p)}{\sqrt{4\pi(\pi n-\delta(p))}}\biggr)\, ,
\en
where $\delta(p)$ denotes the scattering phase, $N$ is the number of
energy levels analyzed and $C$ denotes the normalization constant. 
Below we restrict ourselves to the analysis of the lowest state, putting $N=1$.

In  case of the wide resonance like $\Delta$, it is convenient to consider
the so-called subtracted probability distribution, which is obtained from 
$W(p)$ by subtracting the background $W_{free}(p)$ 
corresponding to the free $\pi N$ pairs
with $\delta(p)=0$~\cite{delta2}. In the vicinity of the resonance, 
the subtracted distribution approximately 
follows the Breit-Wigner form of the scattering 
cross section and thus allows one to easily identify the resonance from
the data on the energy spectrum.

\begin{figure}[t]
\begin{center}
\includegraphics[width=10cm]{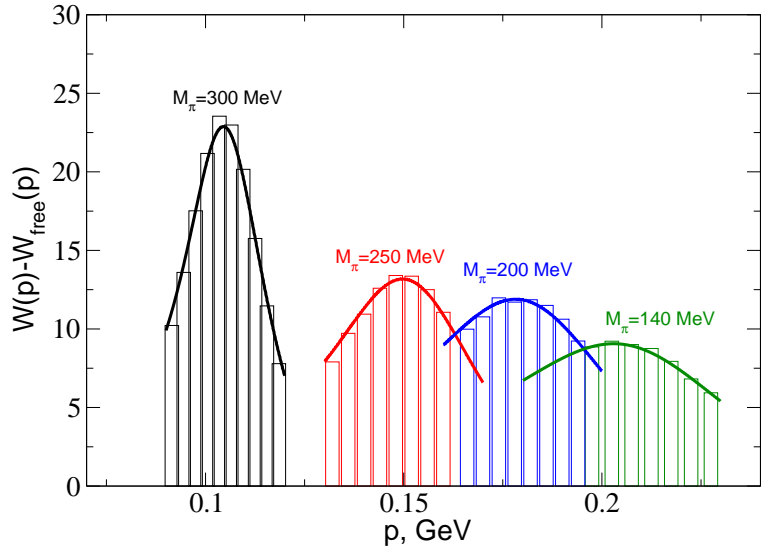}
\end{center}
\caption{Subtracted probability distributions for different values of the pion
mass. The quantity $p$ is the relative momentum of the $\pi N$ pair 
in the center-of-mass frame. The solid lines correspond to the theoretical
prediction based on L\"uscher's formula, see Eq.~(\ref{eq:Wp}).}
\label{fig:distplot}
\end{figure}

Using the values of the various LECs determined from the fit, and substituting
the scattering phase given by Eq.~(\ref{tandelta}) into Eq.~(\ref{eq:Wp}),
one may easily predict the shape of
the probability distributions at different values of the pion
mass. The results are given in Fig.~\ref{fig:distplot}. It is seen that the
distributions behave in the expected manner: for the higher pion
masses, the center-of-mass momentum decreases and the distribution becomes
narrower. Slightly after $300~\mbox{MeV}$ the distribution degenerates
into the $\delta$-function -- the $\Delta$ resonance becomes stable. Of course,
such a behavior can be only observed in practice, 
provided there are at least few 
data points with different values of $L$ at a given quark mass~\cite{delta2}.

\section{Conclusions}
\label{sec:concl}

Here, we summarize the pertinent results of our study:

\begin{itemize}

\item[i)]
We have calculated the ground state energy of the 
$\Delta$ resonance in a finite volume up-to-and-including $O(\epsilon^4)$
in the small scale expansion.

\item[ii)]
The obtained explicit expressions have been used to analyze the recent data 
on the nucleon and $\Delta$ spectrum, provided by the ETM 
collaboration~\cite{Alexandrou}. It turns out that the finite volume
corrections are sizable using the central value for the data point with the smallest quark mass.
Even at the present accuracy, this correction should be taken into account.

\item[iii)]
It is checked that
the numerical values for the correction at $O(\epsilon^3)$ and 
at $O(\epsilon^4)$ do not differ significantly.

\item[iv)]
We perform a simultaneous fit to the nucleon and $\Delta$ masses in the 
infinite volume. The values of the LECs obtained in a result of such a fit
are stable. However, the convergence of the chiral expansion of the baryon 
masses in the infinite volume is rather poor,  since we are
still at relatively high values of the quark masses. 

\item[v)]
The measurement of the energy spectrum at different volumes opens the 
possibility for the extraction of the decay width. To this end, 
we have proposed a procedure based on the consistency 
condition~Eq.~(\ref{eq:condition}).
This procedure can be used meaningfully, 
provided more accurate data points emerge below threshold.

\end{itemize}

\bigskip

\noindent{\em Acknowledgments:}

\bigskip

We would like to thank C.~Alexandrou, Z. Fodor, J.~Gasser, K.~Jansen,
Ch.~Lang, J.~Negele, V.~Pascalutsa, O.~P\`ene, M.~Procura, G.~Schierholz,
M.~Vanderhaeghen and U.~Wenger for interesting discussions.
We acknowledge the support of the European Community-Research Infrastructure 
Integrating Activity ``Study of Strongly Interacting Matter'' 
(acronym HadronPhysics2, Grant Agreement n. 227431)
under the Seventh Framework Programme of the EU.
Work supported in part by DFG (SFB/TR 16, ``Subnuclear Structure of Matter''), 
and  by the Helmholtz Association through funds provided to the virtual institute 
``Spin and strong QCD'' (VH-VI-231).


\renewcommand{\thefigure}{\thesection.\arabic{figure}}
\renewcommand{\thetable}{\thesection.\arabic{table}}
\renewcommand{\theequation}{\thesection.\arabic{equation}}

\vspace{1cm}

\appendix

\setcounter{equation}{0}
\setcounter{figure}{0}
\setcounter{table}{0}

\section{The masses of the nucleon and the $\Delta (1232)$}
\label{app:masses}

A straightforward calculation of the nucleon and $\Delta$ masses at fourth
order yields
\eq
M_N&=&m_N-\frac{3g_A^2M^3}{32\pi F^2}
-\frac{3g_A^2M^4}{64\pi^2F^2m_N}\,\biggl(2\ln\frac{M}{m_N}+1\biggr)
+\frac{3M^4c_2}{128\pi^2F^2}
\nonumber\\[2mm]
&+&\frac{3M^4(8c_1-c_2-4c_3)}{32\pi^2F^2}\,\ln\frac{M}{m_N}
+\frac{M^4}{48\pi^2F^2m_Nm_\Delta^2}\,
\biggl\{P_1+2 P_2\ln\frac{M}{m_N}\biggr\}
\nonumber\\[2mm]
&-&Z\,\frac{((m_\Delta+m_N)^2-M^2)\lambda(m_\Delta^2,m_N^2,M^2)}
{6m_NF^2m_\Delta^2}\,
W_\Delta^r(m_N^2)
\nonumber\\[2mm]
&-&Z\,\frac{(m_\Delta-m_N)(m_\Delta+m_N)^3}{96\pi^2F^2m_Nm_\Delta^2}
\biggl\{\frac{(m_\Delta^2-m_N^2)^2}{6m_N^2}
-2M^2\ln\frac{M}{m_N}
\nonumber\\[2mm]
&-&\frac{M^2(2m_\Delta^2+2m_N^2-m_\Delta m_N)}{3m_N^2}
\biggr\}
+O(\epsilon^5)\, ,
\en
\eq
M_\Delta&=&m_\Delta-\frac{5g_1^2M^3}{96\pi F^2}
-\frac{5g_1^2M^4}{192\pi^2F^2m_\Delta}\,
\biggl(\frac{20}{9}\,\ln\frac{M}{m_N}+\frac{49}{54}\biggr)
+\frac{3M^4a_2}{128\pi^2F^2}
\nonumber\\[2mm]
&+&\frac{3M^4(8a_1-a_2-4a_3)}{64\pi^2F^2}\,2\ln\frac{M}{m_N}
+\frac{M^4}{768\pi^2F^2m_\Delta^5}\,\biggl\{Q_1+2Q_2\ln\frac{M}{m_N}\biggr\}
\nonumber\\[2mm]
&-&Z\,\frac{((m_\Delta+m_N)^2-M^2)\lambda(m_\Delta^2,m_N^2,M^2)}
{24m_\Delta^3F^2}\,W_N^r(m_\Delta^2)
\nonumber\\[2mm]
&-&Z\,\frac{(m_\Delta-m_N)(m_N+m_\Delta)^3}{384\pi^2F^2m_\Delta^3}\,
\biggl\{\frac{(m_\Delta^2-m_N^2)^2}{3m_\Delta^2}
+2M^2\ln\frac{M}{m_N}
\nonumber\\[2mm]
&-&\frac{2M^2(2m_\Delta^2+2m_N^2-m_\Delta m_N)}{3m_\Delta^2}
\biggr\}
+O(\epsilon^5)\, , 
\en
where $M^2=2B\hat m$ and the ``tree-level masses'' are given by
\eq\label{eq:A3}
m_N&=& \mathring{m}_N-4c_1M^2-4B_{23}\Delta_0M^2-B_{32}\Delta_0^3
-E_1\Delta_0^4-4E_2\Delta_0^2M^2-4e_1M^4
\nonumber\\[2mm]
&=& \hat{\mathring{m}}_N-4\hat c_1M^2-4 e_1 M^4\, ,
\nonumber\\[2mm]
m_\Delta&=&\mathring{m}_\Delta-4a_1M^2-4B_1^\Delta\Delta_0M^2-B_0^\Delta\Delta_0^3
-E_1^\Delta\Delta_0^4-4E_2^\Delta\Delta_0^2M^2-4e_1^\Delta M^4
\nonumber\\[2mm]
&=& \hat{\mathring {m}}_\Delta-4\hat a_1 M^2 -4 e_1^\Delta M^4\, ,
\en
where $e_1=4e_{38}+\frac{1}{2}\,(e_{115}+e_{116})$ and
$e_1^\Delta=4e_{38}^\Delta+\frac{1}{2}\,(e_{115}^\Delta+e_{116}^\Delta)$.
Furthermore,
\eq\label{eq:defZ}
Z&=&c_A^2 +2(m_\Delta-m_N)c_Ab_3
+\frac{m_\Delta^2-m_N^2-M^2}{m_N}\,c_Ab_6
\nonumber\\[2mm]
&=&c_A^2+2\Delta_0c_A(b_3+b_6)+O(\epsilon^2)\, ,
\en
and
\eq
P_1&=&\frac{m_N+m_\Delta}{2m_N^2}\,
\biggl\{\frac{c_A^2}{3}(-3m_\Delta^3-3m_\Delta m_N^2+8m_N^3)
\nonumber\\[2mm]
&-&(3m_N^4+2m_\Delta(m_\Delta^2+m_N^2)(m_\Delta-m_N))
c_A\biggl(b_3+\frac{m_N+m_\Delta}{2m_N}\,b_6\biggr)\biggr\}\, ,
\nonumber\\[4mm]
P_2&=&c_A^2(-m_\Delta^2+m_N^2+3m_\Delta m_N)
\nonumber\\[2mm]
&-&(m_\Delta+m_N)(3m_N^2+2m_\Delta^2-2m_\Delta m_N)
c_A\biggl(b_3+\frac{m_N+m_\Delta}{2m_N}\,b_6\biggr)\, ,
\nonumber\\[4mm]
Q_1&=&
c_A^2(3m_\Delta^4+4m_N^4+4m_N^3m_\Delta+4m_N^2m_\Delta^2+4m_Nm_\Delta^3)
\nonumber\\[4mm]
&-&(m_N+m_\Delta)\biggl(2c_Ab_3+\frac{(m_\Delta+m_N)c_Ab_6}{m_N}\biggr)
\nonumber\\[2mm]
&\times&(3m_\Delta^4+4m_N(m_N^2+m_\Delta^2)(m_N-m_\Delta))\, ,
\nonumber
\en
\eq
Q_2&=&-2m_\Delta^2\biggl\{c_A^2(2m_Nm_\Delta+3m_\Delta^2+2m_N^2)
\nonumber\\[2mm]
&-&(m_N+m_\Delta)\biggl(2c_Ab_3+\frac{(m_\Delta+m_N)c_Ab_6}{m_N}\biggr)
(-2m_Nm_\Delta+3m_\Delta^2+2m_N^2)\biggr\}\, .
\nonumber\\
&&
\en
The loop functions are given by
\eq
W_\Delta^r(m_N^2)&=&
\left\{
\begin{array}{l l}
-\frac{\sqrt{-\lambda}}{16\pi^2m_N^2}
\arccos\biggl(-\frac{m_N^2-m_\Delta^2+M^2}{2m_NM}\biggr) &
\\[2mm]
-\frac{m_N^2-m_\Delta^2+M^2}{32\pi^2m_N^2}
\,\biggl(2\ln\frac{M}{m_N}-1 \biggr)\, , &\mbox{if $\lambda<0$}
\\[8mm]
-\frac{\sqrt{\lambda}}{32\pi^2m_N^2}\,
\ln\frac{m_N^2+M^2-m_\Delta^2+\sqrt{\lambda}}
{m_N^2+M^2-m_\Delta^2-\sqrt{\lambda}} &
\\[2mm]
-\frac{m_N^2-m_\Delta^2+M^2}{32\pi^2m_N^2}
\,\biggl(2\ln\frac{M}{m_N}-1 \biggr)\, ,&\mbox{if $\lambda>0$}
\\
\end{array}
\right.\, ,
\nonumber\\[2mm]
W_N^r(m_\Delta^2)&=&
\left\{
\begin{array}{l l}
-\frac{\sqrt{-\lambda}}{16\pi^2m_\Delta^2}
\arccos\biggl(-\frac{m_\Delta^2-m_N^2+M^2}{2m_\Delta M}\biggr) &
\\[2mm]
-\frac{m_\Delta^2-m_N^2+M^2}{32\pi^2m_\Delta^2}
\,\biggl(2\ln\frac{M}{m_N}-1 \biggr)\, , & \mbox{if $\lambda<0$}
\\[8mm]
-\frac{\sqrt{\lambda}}{32\pi^2m_\Delta^2}\,
\ln\frac{m_\Delta^2+M^2-m_N^2+\sqrt{\lambda}}
{m_\Delta^2+M^2-m_N^2-\sqrt{\lambda}} &
\\[2mm]
-\frac{m_\Delta^2-m_N^2+M^2}{32\pi^2m_\Delta^2}
\,\biggl(2\ln\frac{M}{m_N}-1 \biggr)\, ,&\mbox{if $\lambda>0$}
\\
\end{array}
\right. \, ,
\en
where $\lambda=\lambda(m_\Delta^2,m_N^2,M^2)$.

We further express the quantity $M^2$ through the pion mass, according
to
\eq
M^2&=&M_\pi^2\biggl\{1+\frac{M_\pi^2}{32\pi^2F^2}\,
\biggl(\bar l_3+\ln\frac{\bar M_\pi^2}{M_\pi^2}\biggr)\biggr\}\, ,
\en
where $\bar l_3= 2.9\pm 2.4$ is the $O(p^4)$ LEC in the  meson sector of
chiral perturbation theory and $\bar M_\pi$ stands for the physical pion mass.

Finally, normalizing $M_N$ and $M_\Delta$ at $M_\pi=\bar M_\pi$ and
neglecting higher-order terms in the $\epsilon$-expansion, we obtain the 
equations~(\ref{eq:fitformulae}) from section~\ref{sec:infinite},
where

\eq
x_1&=&-4\hat c_1\, ,
\nonumber\\[2mm]
y_1&=&-4\hat a_1\, ,
\nonumber\\[2mm]
x_2&=&-\frac{3g_A^2}{32\pi F^2}\, ,
\nonumber\\[2mm]
y_2&=&-\frac{5g_1^2}{96\pi F^2}\, ,
\nonumber
\en
\eq
x_3&=&-4e_1-\frac{3g_A^2}{64\pi^2F^2\bar m_N}+\frac{3c_2}{128\pi^2F^2}
+\frac{1}{24\pi^2F^2\bar m_N}\,\biggl(\frac{c_A^2}{3}
-\frac{3}{2}\,c_A\bar m_N(b_3+b_6)\biggr)
\nonumber\\[2mm]
&-&\frac{c_1}{8\pi^2F^2}\,\biggl(\bar l_3+\ln\frac{\bar M_\pi^2}{\bar m_N^2}\biggr)\, ,
\nonumber
\en
\eq
y_3&=&-4e_1^\Delta-\frac{5g_1^2}{192\pi^2 F^2 \bar m_\Delta}\cdot\frac{49}{54}
+\frac{3a_2}{128\pi^2F^2}
\nonumber\\[2mm]
&+&\frac{1}{768\pi^2 F^2\bar m_\Delta}\,
\biggl(19c_A^2-12c_A\bar m_\Delta(b_3+b_6)\biggr)
-\frac{a_1}{8\pi^2F^2}\,
\biggl(\bar l_3+\ln\frac{\bar M_\pi^2}{\bar m_N^2}\biggr)\, ,
\nonumber\\[2mm]
x_4&=&-\frac{3g_A^2}{32\pi^2F^2\bar m_N}+\frac{3(8c_1-c_2-4c_3)}{32\pi^2F^2}
+\frac{1}{8\pi^2F^2\bar m_N}\,\biggl(c_A^2-2c_A\bar m_N(b_3+b_6)\biggr)
\nonumber\\[2mm]
&+&\frac{c_1}{4\pi^2F^2}\, ,
\nonumber\\[2mm]
y_4&=&-\frac{5g_1^2}{192\pi^2 F^2 \bar m_\Delta}\cdot\frac{20}{9}
+\frac{3(8a_1-a_2-4a_3)}{32\pi^2F^2}
\nonumber\\[2mm]
&-&\frac{1}{192\pi^2F^2\bar m_\Delta}
\biggl(7c_A^2-12c_A\bar m_\Delta(b_3+b_6)\biggr)+\frac{a_1}{4\pi^2F^2}\, ,
\en
and the ``tree-level'' masses $m_N,m_\Delta$ are given by 
Eq.~(\ref{eq:NDmasses}).

Finally, the loop functions in Eq.~(\ref{eq:fitformulae}) are defined as
\eq
\Phi_N(m_N,m_\Delta,M_\pi^2)&=&
\frac{((m_\Delta+m_N)^2-M_\pi^2)
\lambda(m_\Delta^2,m_N^2,M_\pi^2)}{6m_Nm_\Delta^2}\,W_\Delta^r(m_N^2)
\nonumber\\[2mm]
&+&\frac{(m_\Delta-m_N)(m_\Delta+m_N)^3}{96\pi^2m_Nm_\Delta^2}
\biggl\{\frac{(m_\Delta^2-m_N^2)^2}{6m_N^2}
\nonumber\\[2mm]
&-&\frac{M_\pi^2(2m_\Delta^2+2m_N^2-m_\Delta m_N)}{3m_N^2}
-2M_\pi^2\ln\frac{M_\pi}{m_N}\biggr\}\, ,
\nonumber\\[4mm]
\Phi_\Delta(m_N,m_\Delta,M_\pi^2)&=&
\frac{((m_\Delta+m_N)^2-M_\pi^2)\lambda(m_\Delta^2,m_N^2,M_\pi^2)}
{24m_\Delta^3}\,W_N^r(m_\Delta^2)
\nonumber\\[2mm]
&+&\frac{(m_\Delta-m_N)(m_N+m_\Delta)^3}{384\pi^2m_\Delta^3}\,
\biggl\{\frac{(m_\Delta^2-m_N^2)^2}{3m_\Delta^2}
\nonumber\\[2mm]
&-&\frac{2M_\pi^2(2m_\Delta^2+2m_N^2-m_\Delta m_N)}{3m_\Delta^2}
+2M_\pi^2\ln\frac{M_\pi}{m_N}\biggr\}
\en

\end{document}